\documentclass[showkeys,superscriptaddress,amsmath,amssymb,aps,prl,twocolumn]{revtex4-1}
\usepackage[utf8x]{inputenc}
\usepackage{graphicx}
\usepackage{dcolumn}
\usepackage{bm}
\usepackage{xcolor}
\usepackage{amsmath}
\usepackage{chngcntr}
\usepackage{nicefrac}
\usepackage{booktabs}
\usepackage{blindtext}
\usepackage[colorlinks=true, linkcolor=blue, citecolor=green, urlcolor=cyan]{hyperref}

\usepackage{array}
\newcommand{\PreserveBackslash}[1]{\let\temp=\\#1\let\\=\temp}
\newcolumntype{C}[1]{>{\PreserveBackslash\centering}p{#1}}
\newcolumntype{R}[1]{>{\PreserveBackslash\raggedleft}p{#1}}
\newcolumntype{L}[1]{>{\PreserveBackslash\raggedright}p{#1}}

\begin{document}
\title{Photo-induced Dynamics and Momentum Distribution of Chiral Charge Density Waves in 1T-TiSe$_{2}$}

\author{Qingzheng Qiu}
\affiliation{International Center for Quantum Materials, School of Physics, Peking University, Beijing 100871, China}

\author{Sae Hwan Chun}
\author{Jaeku Park}
\author{Dogeun Jang}
\affiliation{PAL-XFEL, Pohang Accelerator Laboratory, POSTECH, Pohang, Gyeongbuk, 37673 Republic of Korea}

\author{Li Yue}
\affiliation{International Center for Quantum Materials, School of Physics, Peking University, Beijing 100871, China}

\author{Yeongkwan Kim}
\author{Yeojin Ahn}
\author{Mingi Jho}
\author{Kimoon Han}
\affiliation{Korea Advanced Institute of Science and Technology, Daejeon 34141, Republic of Korea}

\author{Xinyi Jiang}
\author{Qian Xiao}
\author{Tao Dong}
\affiliation{International Center for Quantum Materials, School of Physics, Peking University, Beijing 100871, China}
\author{Jia-Yi Ji}
\affiliation{State Key Laboratory of Low Dimensional Quantum Physics and Department of Physics, Tsinghua University, Beijing 100084, China}

\author{Nanlin Wang}
\affiliation{International Center for Quantum Materials, School of Physics, Peking University, Beijing 100871, China}
\affiliation{Collaborative Innovation Center of Quantum Matter, Beijing 100871, China}

\author{Jeroen van den Brink}
\email{j.van.den.brink@ifw-dresden.de}
\affiliation{Institute for Theoretical Solid State Physics, IFW Dresden, Helmholtzstr. 20, 01069 Dresden, Germany}
\affiliation{Würzburg-Dresden Cluster of Excellence ct.qmat, TU Dresden, 01069 Dresden, Germany}
\affiliation{Institute for Theoretical Physics Amsterdam, University of Amsterdam, Science Park904, 1098 XH Amsterdam, The Netherlands}

\author{Jasper van Wezel}
\email{vanwezel@uva.nl}
\affiliation{Institute for Theoretical Physics Amsterdam, University of Amsterdam, Science Park904, 1098 XH Amsterdam, The Netherlands}

\author{Yingying Peng }
\email{yingying.peng@pku.edu.cn}
\affiliation{International Center for Quantum Materials, School of Physics, Peking University, Beijing 100871, China}
\affiliation{Collaborative Innovation Center of Quantum Matter, Beijing 100871, China}

\date{\today}

\begin{abstract}

Exploring the photoinduced dynamics of chiral states offers promising avenues for advanced control of condensed matter systems. Photoinduced or photoenhanced chirality in 1T-TiSe$_{2}$ has been suggested as a fascinating platform for optical manipulation of chiral states. However, the mechanisms underlying chirality training and its interplay with the charge density wave (CDW) phase remain elusive. Here, we use time-resolved X-ray diffraction (tr-XRD) with circularly polarized pump lasers to probe the photoinduced dynamics of chirality in 1T-TiSe$_{2}$. We observe a notable ($\sim$20\%) difference in CDW intensity suppression between left- and right-circularly polarized pumps. Additionally, we reveal momentum-resolved circular dichroism arising from domains of different chirality, providing a direct link between CDW and chirality. An immediate increase in CDW correlation length upon laser pumping is detected, suggesting the photoinduced expansion of chiral domains. These results both advance the potential of light-driven chirality by elucidating the mechanism driving chirality manipulation in TiSe$_2$, and they demonstrate that tr-XRD with circularly polarized pumps is an effective tool for chirality detection in condensed matter systems.
\end{abstract}

\maketitle

In recent years, light-induced phenomena have opened up new possibilities for controlling and manipulating electronic states in condensed matter systems \cite{bao2022light}. A particularly intriguing example is the manipulation of chirality through optical means, which provides a novel approach to studying non-equilibrium chiral dynamics. Chirality, an asymmetry in objects that cannot be superimposed onto their mirror images, is prevalent in natural and synthetic materials \cite{wagniere2007chirality}. In condensed matter systems, chirality is commonly observed in low-symmetry chiral lattices—such as the lattice-locked chiral charge density wave in CoSi \cite{li2022chirality} and the proposed chiral spin density waves in magic-angle twisted bilayer graphene \cite{liu2018chiral}. Notably, the strong coupling between chirality and the lattice necessitates the use of Terahertz-frequency light sources to resonantly excite chiral phonons to induce chiral structure\cite{romao2024phonon,zeng2024breaking}. On the other hand, the spontaneous emergence of chirality from an orbital instability in achiral crystals opens up new frontiers for this topic \cite{van2011chirality,hosur2013kerr,fu2015parity,Qian2024Observation}. The spontaneous gyrotropic order found in 1T-TiSe$_{2}$ in particular, makes it a fascinating potential platform for chirality manipulation \cite{xu2020spontaneous,ishioka2010chiral,castellan2013chiral}.

\begin{figure*}[htbp]
\centering
\includegraphics[width=\textwidth]{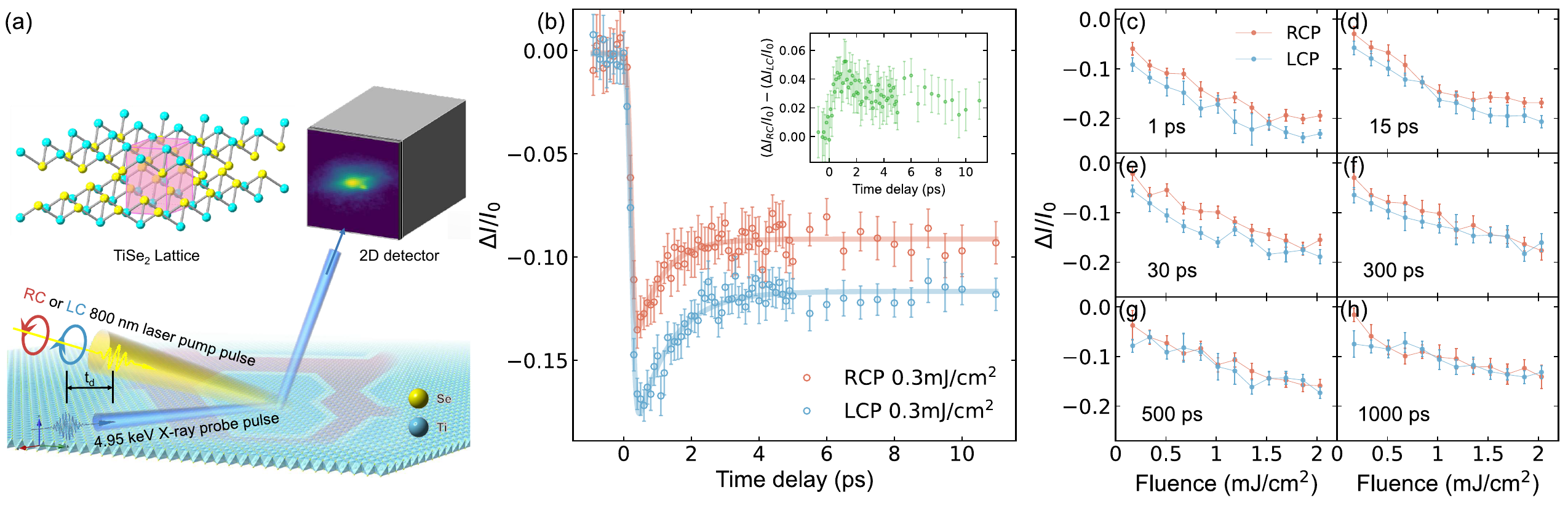}
\caption{{\bf Circular dichroism with transient CDW dynamics after pump.} \textbf{a} Schematic of the ultrafast X-ray diffraction setup, with 800\,nm near-infrared laser as the pump source and Ti-K edge 4.96\,keV X-ray as a probe. The polarization of the laser can be changed by adjusting a quarter-wave plate. \textbf{b} Time evolution of ($\frac{1}{2}$, $\frac{1}{2}$, $\frac{1}{2}$) CDW peak intensity after excited by 800\,nm right circularly polarized (RCP) laser and left circularly polarized (LCP) laser at T = 120\,K. Colored lines illustrate the fit to a single exponential function defined by Equation 1 in the main text. The green dots in the inset represent the value of variation differences between RCP and LCP pump at different delays. \textbf{c-h} Fluence-dependence of CDW peak intensity under RCP and LCP pumping at fixed delay time. Intensity variation are normalized by the equilibrium spectral intensities. Error bars represent the uncertainties in the intensity variation in the measurements.
\label{fig1}}
\end{figure*}

Since the first evidence of a chiral CDW in 1T-TiSe$_{2}$ was provided by scanning tunneling microscope (STM) experiments \cite{ishioka2010chiral}, controversy remains due to different results from STM and XRD measurements \cite{lin2019comment,hildebrand2018local,ueda2021correlation}. Rece
ntly, the possibility to induce a chiral state in TiSe$_{2}$ using optical training was revealed by the observation of the circular photogalvanic effect (CPGE)  \cite{xu2020spontaneous,jog2023optically}. Circularly polarized light was shown to yield different photogalvanic current responses in samples with different preferential chiral domains, induced by illuminating them with high-intensity circularly polarized light while cooling through the CDW transition temperature. Despite these recent advances, the microscopic mechanisms underlying the observed effects of circularly polarized light on the establishment and manipulation of chirality in TiSe$_2$ remain unknown, and the lack of momentum-space resolution precludes establishing a direct correlation between the optically chiral CPGE signals and the long-range three-dimensional CDW in TiSe$_{2}$. To address these issues, probes must be capable of directly accessing time-domain signals while also possessing momentum-space resolution.

Ultrafast spectroscopic studies have demonstrated remarkable capability in elucidating the electronic and structure dynamics in TiSe$_2$ \cite{mohr2011nonthermal,burian2021structural,rohwer2011collapse,porer2014non,mathias2016self,hellmann2012time,lian2020ultrafast,duan2021optical,cheng2022light,cheng2022light}. Time-resolved X-ray diffraction (tr-XRD) is an ideal technique for investigating the dynamic behavior of CDW compounds, such as K$_{0.3}$MoO$_{3}$ \cite{huber2014coherent}, 1T-TaS$_{2}$ \cite{laulhe2017ultrafast,laulhe2015x} and 1T-TiSe$_{2}$ \cite{mohr2011nonthermal,burian2021structural}. Previous experiments predominantly used linearly polarized light as the pump source, despite circularly polarized light being proposed and demonstrated to exhibit high selectivity towards chiral systems \cite{nie2023unraveling,xu2020spontaneous,jog2023optically,tang2010optical}. Here, we investigate the response of the CDW phase in TiSe$_{2}$ under circularly polarized laser pumps and track the dynamic changes of CDW domains. We measured CDW diffraction peaks using Ti-K edge 4.96\,keV X-ray pulses in a grazing incidence geometry with an incidence angle of approximately 1$^\circ$. The pronounced circular dichroism (CD) in the ultrafast CDW dynamics under different circularly polarized pumps was observed. Through the peak profile analysis, we demonstrate that the correlation length of the CDW experiences an instantaneous increase upon circularly polarized laser pumping. These results enrich our understanding of photoinduced dynamics and reveal the mechanisms underlying chiral training under continuous light exposure.

\emph{Circular dichroism in ultrafast dynamics} 1T-TiSe$_{2}$ is a layered transition metal dichalcogenide (TMD) material characterized by van der Waals-coupled adjacent planes. It exhibits a high-temperature symmetry group of $p\overline{3}m1$ and undergoes a commensurate 2$\times$2$\times$2 CDW transition at T$_{c}$ $\sim$ 200\,K \cite{di1976electronic}. The experimental setup is depicted in Fig.~\ref{fig1}a. The chirality within TiSe$_{2}$ was characterized by the transient intensity of the ($\frac{1}{2}$, $\frac{1}{2}$, $\frac{1}{2}$) CDW diffraction peak under right circularly polarized (RCP) and left circularly polarized (LCP) laser pump, as illustrated in Fig.~\ref{fig1}b. The temporal evolution of CDW intensity is traced by integrating the intensities around the CDW regions at different time delays. Within 100\,fs after the laser pump, the diffraction intensity rapidly decreases, followed by a fast recovery process; the system enters a quasi-equilibrium state subsequently, during which no significant recovery is observed within the maximum delay time. Remarkably, despite identical geometrical configurations and pump fluences, the LCP pump is more effective than the RCP pump in suppressing the CDW peak intensity. The rise time of this difference between two pumps is almost identical to that of the CDW suppression. To quantify the observed CD, we fitted the data using the following function:
\begin{equation}
    \frac{I(t)}{I_{0}}=1-\frac{1}{2}\left(1+\operatorname{erf}\left(\frac{t}{\tau_{d}}\right)\right) \times \left(A_{1} \times e^{-\frac{t}{\tau_{r}}} + A_{2}\right)
\end{equation}

where $\tau_{d}$ denotes the decay time, while $A_{1}$ and $\tau_{r}$ respectively represent the proportion and recovery time of the fast recovery process, and $A_{2}$ denotes the proportion of the slow recovery. The results of the fitting process are presented in Table 1. It is noteworthy that under LCP and RCP pumping, the proportion of the fast process components is nearly equivalent to that of the slow process, $\frac{A_{1,lc}}{A_{1,rc}}\approx \frac{A_{2,lc}}{A_{2,rc}}\approx1.2$. This indicates that the CD emerges immediately upon illumination of circularly polarized pump lasers, so even the fast recovery process carries the CD signals. The rapid re-establishment of CD suggests that its origin can be attributed to differences in the screening effect induced by the transient photogenerated charge carriers \cite{rohwer2011collapse,porer2014non,mathias2016self}, which will be discussed in the subsequent section.

\begin{table}[htbp] 
\centering
\caption{Fitting result}
\label{tab:fitting-result}
\begin{tabular}{lC{2.5cm}C{2.5cm}C{2cm}}
\hline 
Pump & $A_{1}$ & $A_{2}$ & $\tau_{r}$ (ps) \\ 
\hline 
RCP & 0.061 $\pm$ 0.010 & 0.094 $\pm$ 0.003 & 0.61 $\pm$ 0.15 \\
LCP & 0.077 $\pm$ 0.005 & 0.114 $\pm$ 0.003 & 1.01 $\pm$ 0.14 \\ 
\hline
\end{tabular}
\end{table}

Fig.~\ref{fig1}c-h plot the fluence dependence of CDW intensity pumped by LCP and RCP lasers at varying fixed delay times. As the delay time increases, the fluence-dependent behaviors of LCP and RCP pumping overlap around 500 ps, indicating the disappearance of CD.  When the CD effect disappears, the system does not fully relax to the pre-excitation state, as depicted in Fig.~\ref{fig1}g, suggesting that the thermal injection from LC and RC polarized lights is equivalent. Additionally, the results of optical absorption measurement indicate no significant difference between LCP and RCP (Supplementary Notes 2).  Due to the grazing incidence configuration, the incident pump light obtains a linearly polarized component, but this does not affect our conclusions about circular dichroism as detailed in Supplementary Notes 8. 

\begin{figure}[htbp]
\centering
\includegraphics[width=\columnwidth]{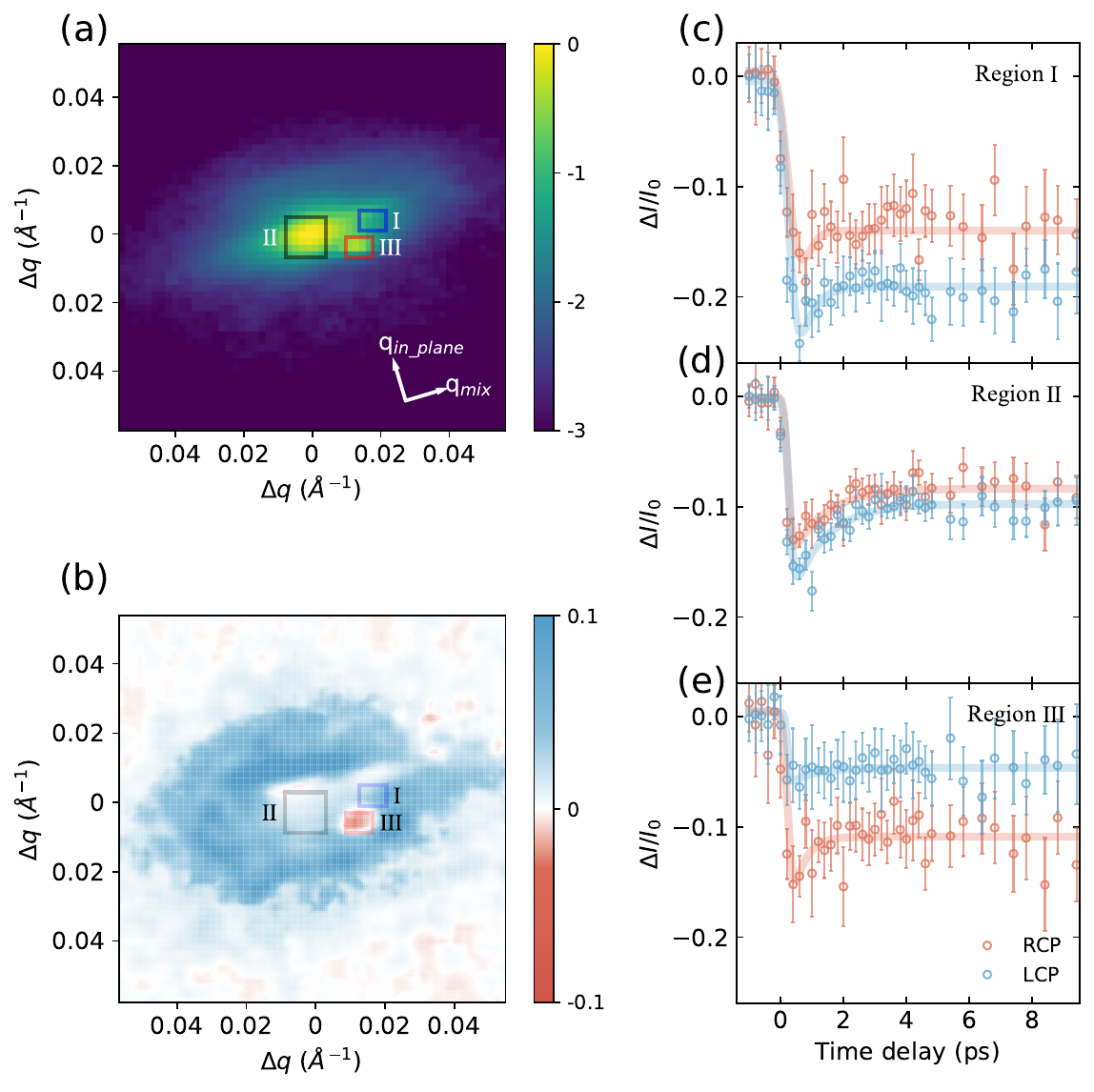}
\caption{{\bf The inhomogeneous distribution of circular dichroism in momentum space.} \textbf{a} The ($\frac{1}{2}$, $\frac{1}{2}$, $\frac{1}{2}$) CDW peak at the equilibrium state in momentum space. The intensities are displayed on a logarithmic scale for clarity. The momentum space can be approximately decomposed into q$_{in\_plane}$ = (0.875$a^*$ + 0.215$b^*$ ) and q$_{mix}$ = (0.157$a^*$ - 0.236$b^*$ +1.916 $c^*$) \textbf{b} The distribution of $\Delta_{CD}$ in momentum space at 1\,ps after laser pump of 800\,nm and 0.3mJ/cm$^2$. The results are computed from the averaged data of multiple scans, which underwent correlation correction to reduce random fluctuations. Data processing details can be found in the main text and supplementary Notes 4. \textbf{c-e} Transient dynamics of integrated intensity within the regions selected in blue, black, and red boxes in (\textbf{a}) and (\textbf{b}), respectively. Same fitting approach as Fig.~\ref{fig1}. Error bars represent the uncertainties in the intensity variation. 
\label{fig2}}
\end{figure}

The origin of CD can be further clarified by examining the distribution of polarization-dependent dynamics in reciprocal space. We define the relative circular dichroism $\Delta_{CD}$(q,t) = (I$_{rc}$(q,t) - I$_{lc}$(q,t)) / (I$_{rc}$(q,t) + I$_{lc}$(q,t)), where I$_{rc}$(q,t) and I$_{lc}$(q,t) denote the spectral intensity in momentum q at delay time t after RCP and LCP pumping, respectively. The distribution of photoinduced CD is presented in Fig.~\ref{fig2}b, uncovering two prominent features: (i) the distribution of circular dichroism in momentum space is inhomogeneous. The positive CD dominates, while Region III shows negative CD (Fig.~\ref{fig2}e). Region III originates from a mosaic peak, which deviates slightly from the main lattice orientation. This mosaic block consists of predominantly opposite chiral domains and leads to an opposite circular dichroism sign from the main lattice. (ii)  The peripheral positions (Region I) displayed a more pronounced relative circular dichroism in comparison to the central peak (Region II), indicating the inhomogeneous ultrafast dynamics. Notice that the results shown here are the average of multiple measurements, which is used to mitigate the effect of random fluctuations; individual measurements reveal even more pronounced spatial inhomogeneity in CD distribution (Supplementary Notes 5).

These features indicate that 1T-TiSe$_{2}$ concurrently exhibits left-handed and right-handed domains, in agreement with STM results \cite{ishioka2010chiral}. These left-handed and right-handed domains exhibit distinct sensitivities to LCP and RCP pumps. In an ideal sample, the extent of domains with either handedness is nearly equal due to their energetic degeneracy. However, external influences like defects and strain may slightly disrupt this degeneracy, leading to a subtle advantage of one-handedness over the other. Notably, the negative CD shown in Fig.~\ref{fig2}e, in contrast to the dominant positive CD, is detected at a mosaic peak arising from another grain. The detection of a grain exhibiting opposing chirality within the same measurement adds credence to the observed phenomenon.

\begin{figure}[htbp]
\centering
\includegraphics[width=\columnwidth]{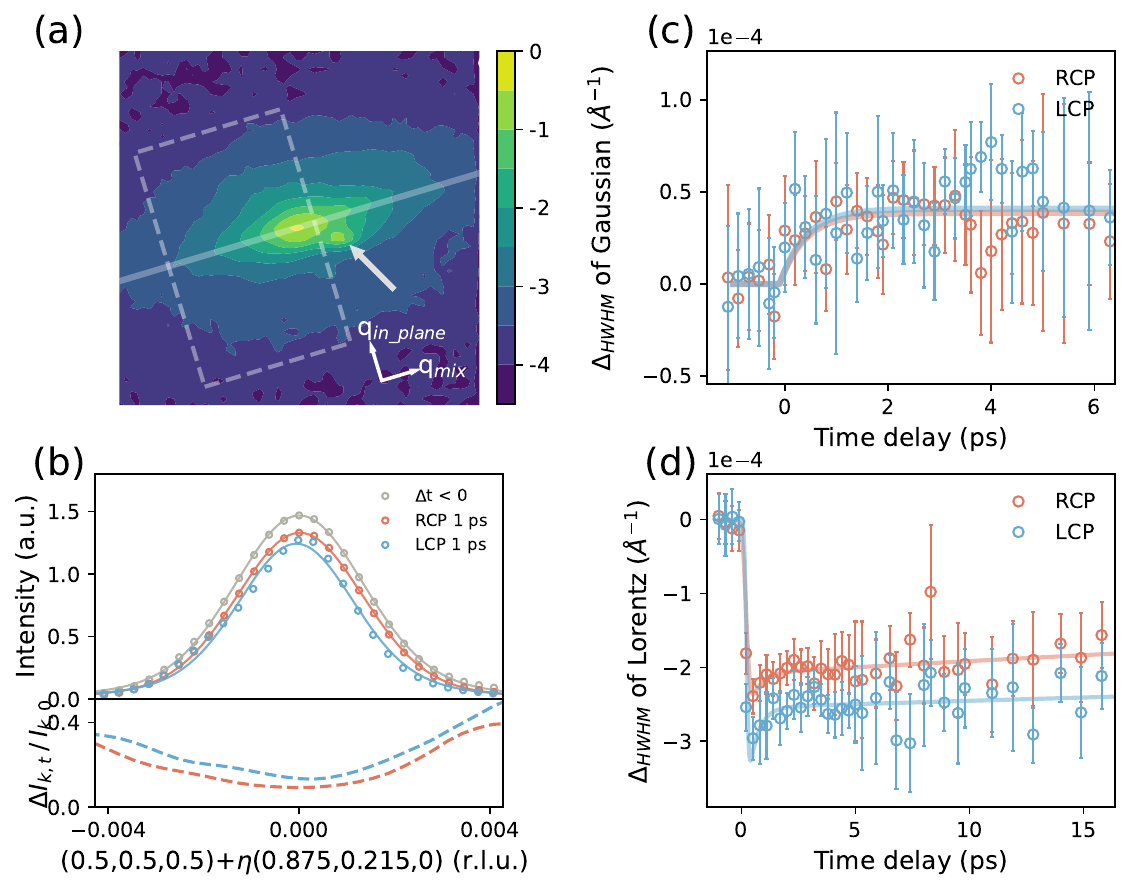}
\caption{{\bf Profile change of CDW peak induced by the circularly polarized laser pump.} \textbf{a} ($\frac{1}{2}$, $\frac{1}{2}$, $\frac{1}{2}$) CDW peak in momentum space, depicted via a contour plot, and a mosaic peak adjacent to the main peak is presented. The white dashed box denotes the integration region selected for extracting the main CDW peak profile in \textbf{(b)}, and the white solid line shows the direction of integration, which is parallel to q$_{mix}$. The white arrow marks the mosaic peak. \textbf{b} In-plane profiles of the CDW peak at equilibrium state and 1 ps after right and left circularly polarized pump, respectively. Dashed lines in the lower panel depict the ratio of the intensity variation ${\Delta}I_{k,t}=I_{k,0}-I_{k,t}$ to its equilibrium value $I_{k,0}$ at respective momentum k, the data presented are obtained by directly dividing the pumped data by the unpumped data without any subtraction. \textbf{c}, \textbf{d} Time dependence of HWHW variations for Gaussian and Lorentz components, respectively, obtained by subtracting the fit parameters before- and after-pump, as extracted from Voigt profile fittings. Lines are for visual guidance. Error bars represent the uncertainties in the Voigt profile fitting.
\label{fig3}}
\end{figure}

\emph{Increase of correlation length}—To demonstrate the impact of circularly polarized pump light on CDW domain walls, an analysis of the CDW peak profile is imperative. As illustrated in Fig.~\ref{fig3}a, a region delineated by the white box was selected to circumvent interference from the mosaic peak. Integration was performed along the white solid line depicted in Fig.~\ref{fig3}a, wherein out-of-plane spectral weights were integrated to extract the in-plane characteristics. The extracted profiles of CDW peak at equilibrium state and 1 ps after the RCP and LCP pump are depicted in Fig.~\ref{fig3}b. The variation in the spectral intensity is non-uniform, with a 10\% change at the center and a more pronounced 40\% change at the edges, suggesting spectral narrowing. To quantify the peak narrowing, Fig.~\ref{fig3}c,d depict the time dependence of the Gaussian and Lorentzian components, respectively, which are extracted from a Voigt fit (Supplementary Notes 6). On the one hand, the Gaussian component in the Voigt profile, which embodies instrument and thermal vibration effects, exhibits a slow and minor increase in half width at half maximum (HWHM) over a relaxation time of approximately 1 ps. This trend likely stems from the lattice temperature increase, as a $\sim$1 ps relaxation time is consistent with the timescale for energy transfer into phonon baths due to electron-phonon interactions in the two-temperature model \cite{anisimov1974emission,porer2014non}. On the other hand, the Lorentzian component, primarily associated with the intrinsic lattice dynamics and crystal defects, exhibits a decrease in HWHM, indicating an increase in the correlation length. The HWHM of the Lorentzian component exhibits a rapid decline within 200 fs, followed by a slower recovery process with a relaxation time of more than 100 ps. It implies a persistently anomalous increase in correlation length after pumping. Surprisingly, the increase in correlation length exhibits circular polarization dependence, and the recovery timescale is close to the disappearance of CD, as shown in Fig.~\ref{fig1}c-h, which implies a correlation between the two phenomena. The slow recovery may involve chiral domain-wall reconfiguration or topological defect dynamics \cite{vogelgesang2018phase,zong2019evidence}, as pure electronic processes are typically faster.

\emph{Discussion}—The differential suppression of CDW peak intensities under LCP and RCP lights reaches 20\%, far exceeding the CD magnitude of TiSe$_2$ detected by conventional optical reflectivity, which is typically below 1\% \cite{wickramaratne2022photoinduced,nie2023unraveling,meyer2013recent}. This highlights the high sensitivity of our measurements. The observed CD effect is not due to artifacts commonly encountered in all-optical pump-probe experiments \cite{meyer2013recent}, as the utilization of the X-ray probe precludes the possibility of direct interference from the pump radiation. The possibility of CD originating from different fluences between two circular pumps can be excluded, as it fails to account for the observed mosaic peaks with opposite chirality and the disappearance of CD at 500 ps, where the system has not yet fully recovered from thermal effects. 

\begin{figure}[htbp]
\centering
\includegraphics[width=\columnwidth]{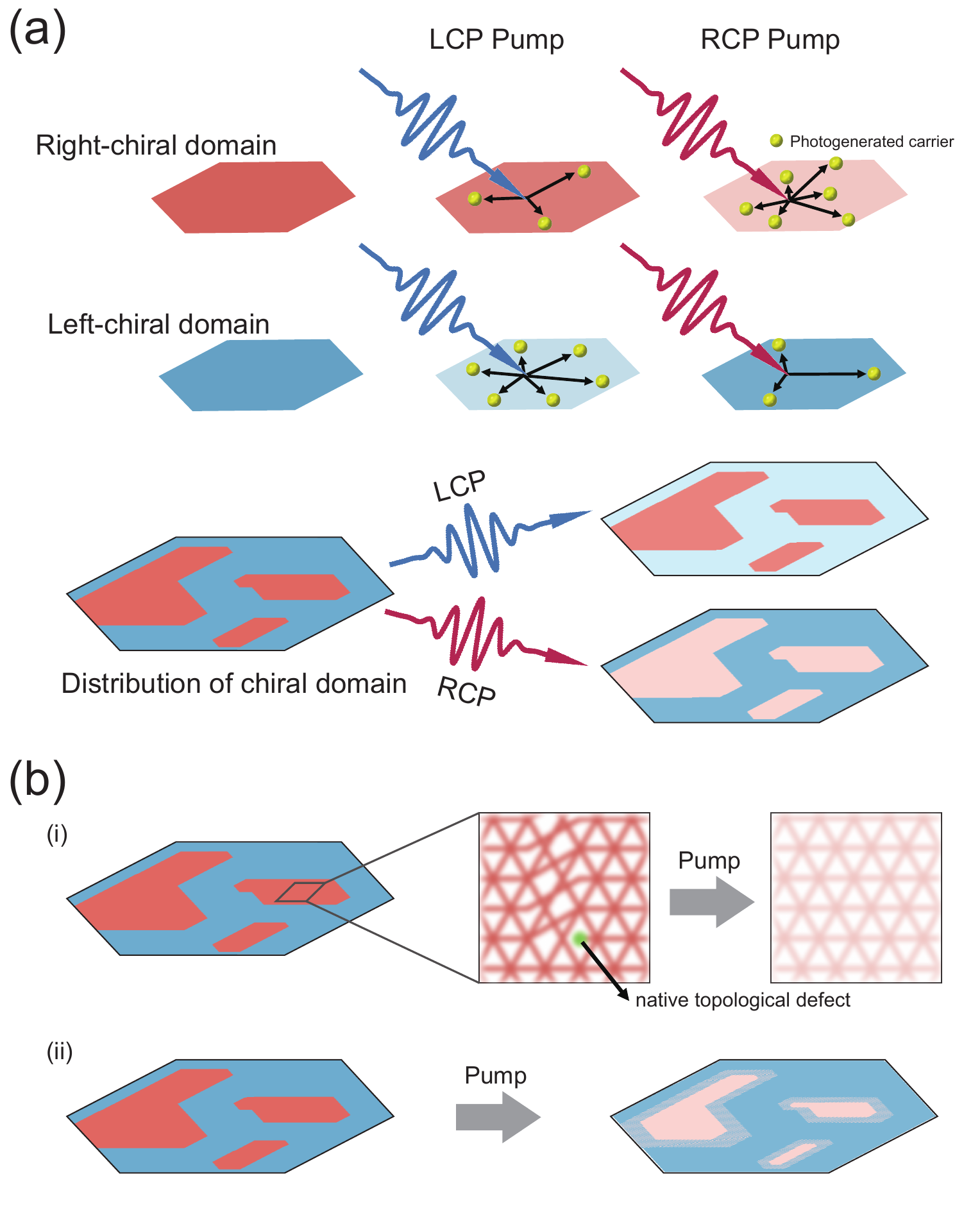}
\caption{{\bf Schematic of different microscopic mechanisms.} \textbf{a} Possible mechanism of the emergence of circular dichroism in CDW dynamics. The red and blue colors are used to indicate domains that are sensitive to LCP and RCP pumps, respectively. The darker shades of color indicate a stronger CDW amplitude, while the lighter shades indicate a weaker CDW amplitude. Circularly polarized pumping may excite a disparate population of carriers, thereby introducing a chiral dependence in the CDW quenching process. \textbf{b} Possible mechanisms for the enhanced coherence length. The stripe's brightness indicates the CDW's amplitude. (i) The pump-induced annihilating effect prompts the migration or destruction of native topological defects, consequently increasing the correlation length. (ii) Circularly polarized pump light induces expansion of the corresponding chiral domains, resulting in an elevation of the overall correlation length.
\label{fig4}}
\end{figure}

Chiral domains, with their varied responses to circularly polarized pumped light, are essential to capture the different facets of the observed CD, as illustrated in Fig.~\ref{fig4}a. Within left-chiral domains, the LCP pump may preferentially excite carriers in symmetry-breaking orbitals, resulting in a higher population of photo dissipaters than its right-handedness counterparts. This phenomenon is analogous to the observation that circularly polarized light yields a higher photocurrent than its counter-helicity in CPGE experiments \cite{xu2020spontaneous}. This disparity in carrier density rapidly manifests within 100 fs, and the enhanced screening effect within left-chiral domains leads to a more pronounced suppression of spectral intensity. 
The underlying mechanism may relate to the differential sensitivity of electronic bands to pumps with different circular polarizations.  
Minor disparities in chiral CDW gaps may evolve into significant distinctions in the carrier population through this self-amplification effect \cite{mathias2016self,lian2020ultrafast}. While the origin of this pronounced CD remains to be fully elucidated, the intimate connection between CDW and chirality is compellingly demonstrated. The capability to probe CDW at finite momentum underscores the specificity of the tr-XRD approach.

In addition, circularly polarized pumps also lead to an increase in its correlation length. To explain this, we note that ultrafast pumping might induce an annealing effect, annihilating topological defects and thereby extending the CDW correlation length, as depicted in Fig.~\ref{fig4}b. The higher photoinduced carrier density in dominant left-chiral domains makes their annealing effect more noticeable, in agreement with Fig.~\ref{fig3}d. Alternatively, it is possible that RCP and LCP pumps promote the expansion of left and right chiral domains, respectively, resulting in an overall increase in the correlation length. Both scenarios align with our findings, and further studies are needed to distinguish between them. In either case, our observation that circularly polarized pumping facilitates the formation of one type of domain (or the suppression of the other) suggests a method to induce chiral asymmetry by breaking the degeneracy between domains of opposite handedness. The observed photoinduced circular dichroism reflects the equilibrium state's inherent asymmetry and suggests that chiral domains can exist at equilibrium.

TiSe$_2$ was found to exhibit enhanced out-of-plane photocurrent under LCP (RCP) light if the samples were continuously illuminated with LCP (RCP) light during cooling \cite{xu2020spontaneous}. Despite previous observations suggesting that 800 nm continuous wave laser would not induce a net chirality \cite{xu2020spontaneous}, recent experiments have detected CPGE signals at 800 nm wavelength \cite{jog2023optically}, which is consistent with our results. The observation of circular dichroism in ref. \cite{jog2023optically} requires the continuous-wave laser power to exceed a threshold of 10 $\mu$W, a condition easily met by our femtosecond laser's power.  While direct comparison with previous experiments necessitates caution due to the various distinctions between femtosecond laser and continuous wave laser, our results may elucidate the mechanisms underlying chiral training in ref.\cite{xu2020spontaneous}. Exposure to circularly polarized light favors the formation of chiral-sensitive domains by increasing the correlation length. Continuous illumination during the phase transition stabilizes this tendency, resulting in the dominance of one chiral domain, thereby leading to the observed net chirality. In our experiment, the CD lasts for $\sim$500\,ps, whereas the interval between laser pulses is 33\,ms. This interval allows the chirality differentiation to relax before the arrival of the next circularly polarized pulse, preventing permanent overwriting of the global chirality. Harnessing ultrafast lasers with high repetition rates allows for the stabilization and extension of specific chiral domains, enabling real-time control and switching of chirality with selective polarizations, which warrants further investigation.

\emph{Conclusions}-Our work has revealed the connection between CDW and chirality by uncovering the sensitivity of CDW transient dynamics to the helicity of incident lights. These findings suggest the presence of inherent chiral structures within TiSe$_{2}$ and the principles for manipulation via circularly polarized light, wherein ultrafast circularly polarized light pulses amplify the chiral asymmetry in the sample, steering distinct chiral domains into disparate non-thermal evolution pathways. While definitive mechanisms underlying the photoinduced correlation length growth remains elusive, these discoveries provide explanations for chiral phenomena observed in previous experiments \cite{xu2020spontaneous} and offer new insights into the probing and understanding of chirality in condensed matter systems, thus paving a new way to control chirality in materials. Utilizing tr-XRD as an effective tool for detecting chirality introduces a new paradigm for understanding light-matter interactions in chiral electronic systems.

Y.Y.P. is grateful for financial support from the Beijing Natural Science Foundation (Grant No. JQ24001), the Ministry of Science and Technology of China (Grants No. 2024YFA1408702 and No. 2021YFA1401903), and the National Natural Science Foundation of China (Grants No. 12374143 and No. 11974029). J.B. was supported by the Deutsche Forschungsgemeinschaft (DFG, German Research Foundation) through the Sonderforschungsbereich SFB 1143, grant No. YE 232/1-1, and under Germany's Excellence Strategy through the W\"{u}rzburg-Dresden Cluster of Excellence on Complexity and Topology in Quantum Matter -- \emph{ct.qmat} (EXC 2147, project-ids 390858490 and 392019). This tr-XRD experiment was performed at the FXS-endstation of the PAL-XFEL funded by the Korea government (MSIT). Data and network service was supported by GSDC and KREONET by the Korea Institute of Science and Technology Information (KISTI).



\end{document}